\documentclass[sigconf]{acmart}
\AtBeginDocument{%
}

\copyrightyear{2024}
\acmYear{2024}
\setcopyright{rightsretained}
\acmDOI{10.1145/3626772.3657942}

\acmConference[SIGIR '24]{Proceedings of the 47th International ACM SIGIR Conference on Research and Development in Information Retrieval}{July 14--18, 2024}{Washington, DC, USA}
\acmBooktitle{Proceedings of the 47th International ACM SIGIR Conference on Research and Development in Information Retrieval (SIGIR '24), July 14--18, 2024, Washington, DC, USA}
\acmISBN{979-8-4007-0431-4/24/07}

\makeatletter
\gdef\@copyrightpermission{
 \begin{minipage}{0.3\columnwidth}
  \href{https://creativecommons.org/licenses/by/4.0/}{\includegraphics[width=0.90\textwidth]{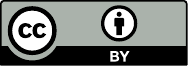}}
 \end{minipage}\hfill
 \begin{minipage}{0.7\columnwidth}
  \href{https://creativecommons.org/licenses/by/4.0/}{This work is licensed under a Creative Commons Attribution International 4.0 License.}
 \end{minipage}
 \vspace{5pt}
}

\makeatother

\definecolor{others}{rgb}{0.154, 0.43, 0.41}
\definecolor{gpttfive}{rgb}{0.253, 0.159, 0.38}
\definecolor{tfive}{rgb}{0.0, 0.116, 0.20}

\usepackage{tabularx}
\usepackage{adjustbox}
\usepackage{multirow}
\usepackage{subfig}
\usepackage{fontawesome}

\begin{document}

\title{Synthetic Test Collections for Retrieval Evaluation}


\author{
Hossein A.~Rahmani
}
\orcid{0000-0002-2779-4942} 
\affiliation{%
        \institution{University College London}
        \city{London}
        \country{UK}
}
\email{hossein.rahmani.22@ucl.ac.uk}

\author{Nick Craswell}
\orcid{0000-0002-9351-8137} 
\affiliation{%
        \institution{Microsoft}
        \city{Bellevue}
        \country{US}
}
\email{nickcr@microsoft.com}

\author{Emine Yilmaz}
\orcid{0000-0003-4734-4532} 
\affiliation{%
        \institution{University College London \& Amazon}
        \city{London}
        \country{UK}
}
\email{emine.yilmaz@ucl.ac.uk}

\author{Bhaskar Mitra}
\orcid{0000-0002-5270-5550} 
\affiliation{%
        \institution{Microsoft}
        \city{Montréal}
        \country{Canada}
}
\email{bmitra@microsoft.com}
\orcid{0000-0002-5138-8426}
\author{Daniel Campos}
\affiliation{%
  \institution{Snowflake}
  \city{New York}
  \country{US}
  }
\email{daniel.campos@snowflake.com}

\renewcommand{\shortauthors}{Hossein A.~Rahmani, Nick Craswell, Emine Yilmaz, Bhaskar Mitra, Daniel Campos}

\begin{abstract}
Test collections play a vital role in evaluation of information retrieval (IR) systems. Obtaining a diverse set of user queries for test collection construction can be challenging, and acquiring relevance judgments, which indicate the appropriateness of retrieved documents to a query, is often costly and resource-intensive. Generating synthetic datasets using Large Language Models (LLMs) has recently gained significant attention in various applications. In IR, while previous work exploited the capabilities of LLMs to generate synthetic queries or documents to augment training data and improve the performance of ranking models, using LLMs for constructing synthetic test collections is relatively unexplored. Previous studies demonstrate that LLMs have the potential to generate synthetic relevance judgments for use in the evaluation of IR systems. In this paper, we comprehensively investigate whether it is possible to use LLMs to construct fully synthetic test collections by generating not only synthetic judgments but also synthetic queries. In particular, we analyse whether it is possible to construct reliable synthetic test collections and the potential risks of bias such test collections may exhibit towards LLM-based models. Our experiments indicate that using LLMs it is possible to construct synthetic test collections that can reliably be used for retrieval evaluation.

\begin{center}
    \faicon{github} \url{https://github.com/rahmanidashti/SyntheticTestCollections}
\end{center}

\end{abstract}





\maketitle

\section{Introduction}
\label{sec:introduction}
Test collection construction is a pivotal process to evaluate the effectiveness of information retrieval (IR) systems. The most widely used approach for constructing test collections is based on the Cranfield paradigm \cite{cleverdon1967cranfield,aslam2006statistical}, which involves creating a collection comprising queries and associated relevance judgments. Queries used in test collection construction are expected to come from real usage logs, representing real information needs. However, it is very difficult to get access to such logs outside of search engine companies. Hence, lots of existing test collections used in IR are based on manually created queries \cite{craswell2021trec,yilmaz2008simple}. This process demands time and expertise, making it costly in terms of labor and resources; furthermore, there are no guarantees that queries generated at the end of this process are representative of real information needs. Similarly, obtaining relevance judgments for a query is an expensive procedure requiring significant human effort. The highly demanding nature of the test construction process is a major bottleneck in constructing large test collections; hence, most existing publicly available test collections tend to consist of a small number of queries, which could degrade the reliability of these collections.

Recently, Large Language Models (LLMs) have demonstrated remarkable performance on unseen tasks by only considering the instructions (so-called `prompts') provided to them \cite{chiang2023can,jiang2022promptmaker}. Synthetic datasets generated using LLMs have recently gained attention across a range of diverse tasks \cite{li2023synthetic,zhang2018synthetic,bao2023synthetic}. In IR, previous studies use LLMs to generate synthetic training data for augmentation to boost the quality of retrievers \cite{askari2023expand,bonifacio2022inpars}. LLMs have also been used to generate relevance labels \cite{thomas2023large,faggioli2023perspectives}, as well as to generate query variants for evaluation and training of IR systems \cite{rajapakse2023improving,alaofi2023can}. However, to the best of our knowledge, no prior study has thoroughly explored the potential of LLMs to construct fully synthetic test collections where both queries and associated relevance judgments are automatically generated using LLMs. 

Given the aforementioned challenges in constructing large-scale test collections, our goal in this paper is to investigate whether it is possible to create reliable synthetic test collections so that there is (i) no need for real usage query logs or manual creation of queries, and (ii) no need to obtain manual relevance judgments. We investigate different approaches to construct synthetic test collections using LLMs and show using synthetic test collections it is possible to obtain evaluation results that are similar to results obtained using real test collections.

\begin{figure}
    \centering
    \includegraphics[scale=0.38]{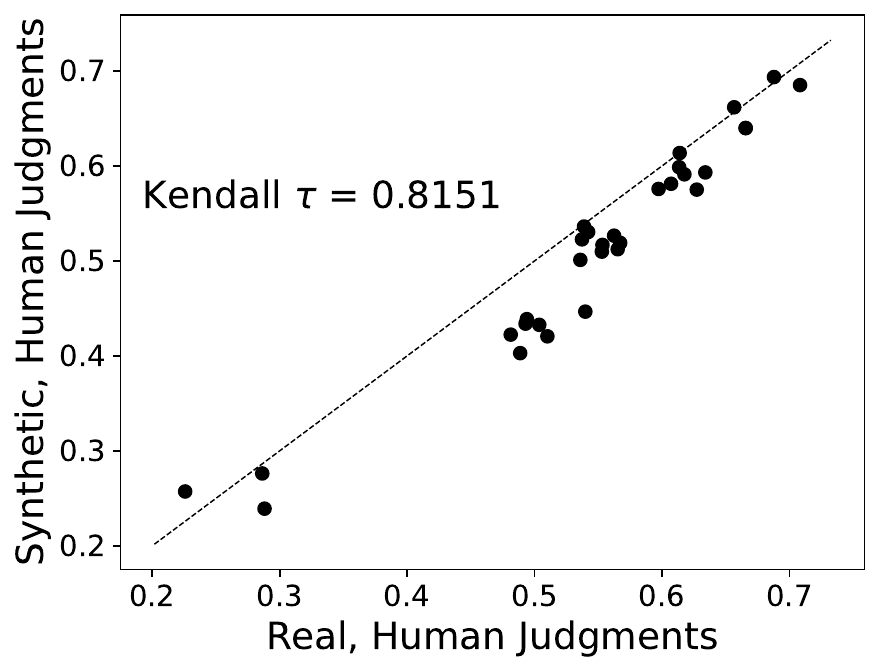}
    \caption{Scatter plot of the effectiveness (i.e., NDCG@10) of TREC DL 2023 runs according to the real queries and synthetic queries with human judgments. A point represents a single run averaged over all queries.}
    \label{fig:realXsynthetic_humanXhuman}
\end{figure}
\section{Synthetic Test Collection}
\label{sec:experiment}

\begin{table}
    \centering
    \caption{Statistics of queries per query type}
    \label{tbl:queries_statistics}
    \begin{adjustbox}{max width=\columnwidth}
        \begin{tabular}{lccccc}
            \toprule
            \textbf{} & \textbf{All} & \textbf{Real} & \textbf{T5 Generated} & \textbf{GPT-4 Generated} \\
            \midrule
             No.~of Queries & 82 & 51 & 13 & 18 \\
             Avg.~Query Length & 6.84 & 5.76 & 5.69 & 10.72 \\
             Min Query Length & 2 & 2 & 4 & 6 \\
             Max Query Length & 15 & 14 & 8 & 15 \\
            \bottomrule
        \end{tabular}
    \end{adjustbox}
\end{table}

\subsection{Synthetic Query Generation}
\label{sec:synthetic_query_generation}
The reliability of a test collection highly depends on the queries included in the collection. Since obtaining real user queries may not always be possible, we analyzed the feasibility of using synthetic queries in test collection construction, focusing on constructing test collections for passage retrieval task. 

Our goal is to create meaningful queries, which can be characterized as a collection of terms that can be used to reach a document and make sense alone without any further context from the document from which the query is derived. The approach we followed is based on starting with a passage and generating a query to which this passage would be relevant. 

To generate the synthetic queries, we first sampled $1000$ passages at random from the MS MARCO v2 passage corpus\footnote{\url{https://msmarco.blob.core.windows.net/msmarcoranking/msmarco_v2_passage.tar}} as anchor documents. The selected passages are then go through a filtering step whose goal is to get rid of ``bad'' passages, which can lead to low quality queries being generated. Examples of bad passages include passages that talk about a person without saying the person's name (``she then completed her first novel\ldots'') or passages that use terms that might not be clear once the passage is removed from its document context (``we eventually chose the second solution, since it gave a good balance of accuracy and efficiency''). To identify passages that might not be good stand-alone search results, we ran each of the $1000$ passages through a GPT-4 prompt, which generates a \textit{query independent passage quality score}. Since this is a rough one-off filter, we did not repeat it in the case of some error or malformed output, which eliminated $8.9\%$ of passages. We filtered low-quality passages with passage quality scores less than $50$, removing $14.6\%$ passages. The remaining passages were seeds for generating synthetic queries.

We then generated queries using two methods: The first method uses a small pre-trained model based on T5 \cite{raffel2020exploring}, and the second method is based on zero-shot query generation using GPT-4 \cite{achiam2023gpt}. For query generation using T5, we applied the BeIR query generation model\footnote{\url{https://huggingface.co/BeIR/query-gen-msmarco-t5-large-v1}} that uses a T5-based model pre-trained on the MS MARCO Passage dataset \cite{thakur2021beir}. We then generated one query per passage using both T5 and GPT-4 approaches. The details of the approaches can be found in our \href{https://github.com/rahmanidashti/SyntheticTestCollections}{GitHub}.

\begin{table}
    \centering
    \caption{Average number of documents per query for each relevance grade for different query types.}
    \label{tbl:numrels_perquery}
    \begin{adjustbox}{max width=\columnwidth}
        \begin{tabular}{lccccc}
            \toprule
            \textbf{Relevance Grade} & \textbf{All} & \textbf{Real} & \textbf{T5 Generated} & \textbf{GPT-4 Generated} \\
            \midrule
             Nonrelevant (0) & 169.09 & 159.31 & 213.30 & 164.88 \\
             Related (1) & 53.31 & 64.15 & 31.23 & 38.55 \\
             Highly relevant (2) & 27.54 & 31.60 & 19.46 & 21.88 \\
             Perfectly relevant (3) & 22.31 & 24.35 & 21.07 & 17.44 \\
             % \hdashline
             % all & 22,327 & 14,251 & 8,076 & 3,706 & 4,370 \\
            \bottomrule
        \end{tabular}
    \end{adjustbox}
\end{table}

Then, we asked experts who are professional assessors with very good experience in relevance annotation to remove the queries that did not look reasonable, that contained too few or too many relevant documents as these queries tend to be noisy or not very informative for evaluation purposes. Amongst $48$ T5-generated queries, $13$ of them were selected and amongst $49$ queries generated using GPT-4, $18$ of them were selected to be included in the test collection. Queries generated as a result of this process were included in the test collections created as part of the TREC Deep Learning Track (TREC DL) 2023 \cite{craswell2024overview}, which consist of $51$ real queries in addition to the synthetic queries.
% \footnote{\url{https://microsoft.github.io/msmarco/TREC-Deep-Learning.html}} 
Systems submitted to the DL Track were run on both synthetic and real queries in the test collection. For all query types, depth-10 pooling was used to select the documents to be judged by the expert assessors who labelled the documents based on four relevance grades (irrelevant, related, highly relevant and perfectly relevant). 

Table~\ref{tbl:queries_statistics} shows the total number of queries included in the TREC DL 2023 test collection for each query type, together with the average, min and max query length for each category. We also included statistics about real queries for a better comparison of synthetic queries with human queries. It can be seen that queries generated using GPT-4 tend to be much longer than the queries generated using T5 and the real queries and in general, queries generated using T5 tend to be shorter than the other query types. For each query type, Table~\ref{tbl:numrels_perquery} shows the average number of documents per query for each relevance grade. It can be seen that synthetic queries contain much fewer relevant documents (documents with a relevance grade greater than zero) compared to the real queries ($120.1$ documents of relevance grade greater than zero for real queries vs.~$71.76$ documents for queries generated by T5 and $77.87$ documents for queries generated by GPT-4.) Also, pools constructed using queries generated by T5 tend to contain significantly more non-relevant documents compared to the rest of the query types, suggesting that these queries may be more difficult than the other queries. 

To compute the reliability of the synthetic queries, we compared the performance of systems that were submitted to TREC DL 2023 on real queries with the performance of systems solely on synthetic queries. For this purpose, we evaluate the quality of the $31$ systems submitted to the full ranking task of the track using official judgments of the track obtained from expert human assessors from NIST, and compare the ranking of these systems on real queries and synthetic queries.
 Figure \ref{fig:realXsynthetic_humanXhuman} shows how the performance of systems using synthetically generated queries (both T5 and GPT-4 based) compare with system performance on real queries. The figure also includes the line $y=x$ for easy comparison.
% Figure \ref{fig:realXsynthetic_humanXsparse} shows the system ranking and Kendall's $\tau$ correlation on real versus synthetic queries on the sparse qrels. 
It can be seen that synthetic queries and real queries show similar patterns in terms of evaluation results and system ranking, with a system ordering agreement of Kendall's $\tau=0.8151$.

\subsection{Synthetic Relevance Judgment Generation}
Once a sample of queries to be included in a test collection has been identified, the second step is to identify documents relevant to these queries, a process that is usually done by collecting manual annotations. While in the previous section we assumed that manual annotations are available for all query types (including synthetic queries), obtaining manual annotations is a very expensive process and it may not always be possible to get these many judgments from expert annotators. Hence, we next investigate the possibility of generating synthetic relevance judgments for test collections constructed using synthetic queries. 

\begin{figure}
    \centering
    \subfloat[\label{fig:realXsynthetic_humanXsparse}]
    {{\includegraphics[scale=0.3]{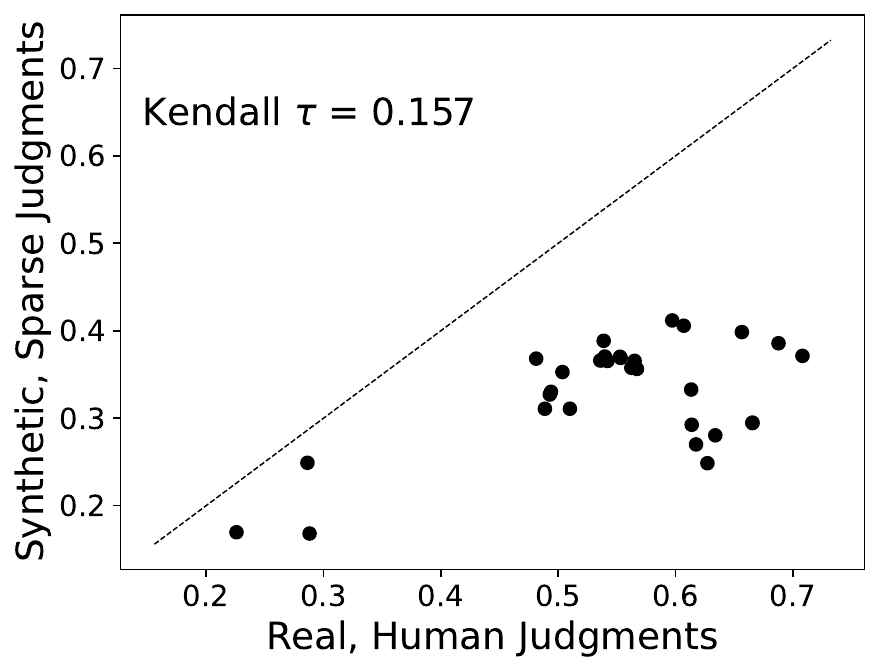} }}%
    % \quad
    \subfloat[\label{fig:realXsynthetic_humanXllm}]
    {{\includegraphics[scale=0.3]{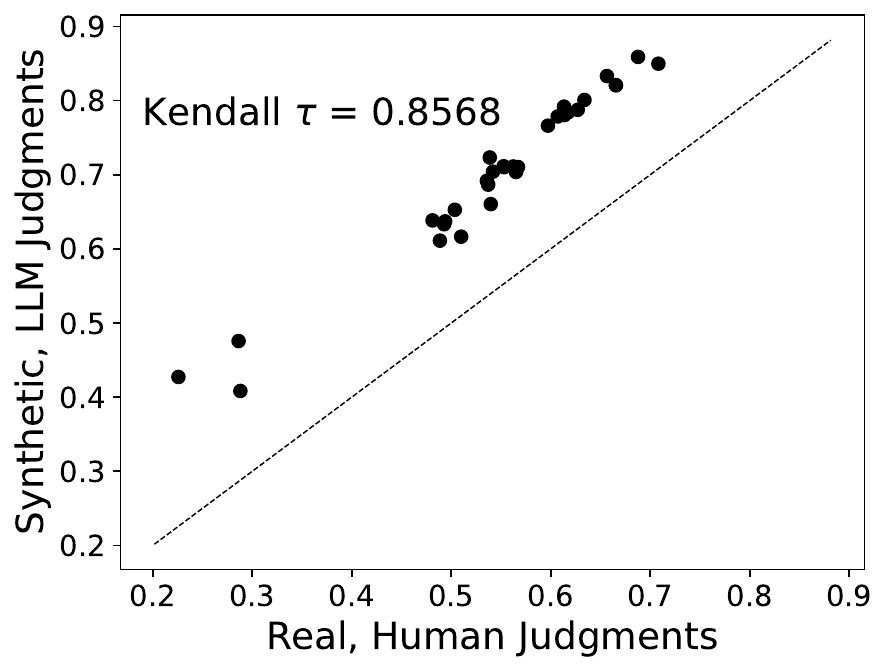} }}%
    \caption{Scatter plots of the effectiveness (i.e., NDCG@10) of TREC DL 2023 run according to the real queries with human judgments and our synthetic queries with (a) sparse judgments and (b) synthetic judgments.}
    \label{fig:sparse_vs_synthetic_labels}%
\end{figure}

As the simplest method, one can assume that the passages that we used to generate the synthetic queries are the only passages that are relevant to those queries, referred to as \textit{sparse judgments}. Figure \ref{fig:realXsynthetic_humanXsparse} shows the system ranking and Kendall's $\tau$ correlation on real queries with human judgments and synthetic queries with sparse judgments. Compared to synthetic queries with human judgments ($\tau = 0.8151$), evaluation using synthetic queries with sparse judgments does not give similar results to human queries, with a very low system ordering agreement of $\tau=0.157$. 

\begin{table}
    \centering
    \caption{Judgment agreement on TREC DL 2023 between TREC assessors and the LLM (i.e., GPT-4) on real queries and synthetic queries.}
    \label{tbl:judgment_agreement_real_queries}
    \begin{adjustbox}{max width=\columnwidth}
        \begin{tabular}{llcccccc}
            \toprule
            \textbf{} & \textbf{GPT-4} & \multicolumn{4}{c}{\textbf{TREC DL 2023 Assessments}} & \multirow{2}{*}{\textbf{Cohen's $\kappa$}}\\
            \cline{3-6}
            \textbf{} & \textbf{Prediction} & \textbf{Perfect rel.} & \textbf{High.~rel.} & \textbf{Related} & \textbf{Irrelevant} & \\
            % \midrule
            % NIST & - & 1830 & 2259 & 4372 & 13866 \\
            \midrule
             \parbox[t]{2mm}{\multirow{4}{*}{\rotatebox[origin=c]{90}{Real}}} & Perfect.~rel.~& \textbf{597} & 469 & 496 & 324 & \multirow{4}{*}{0.24}\\
              & High.~rel.~& 322 & \textbf{473} & 648 & 501 \\
              & Related & 298 & 548 & \textbf{1358} & 2736 \\
              & Irrelevant~& 25 & 122 & 770 & \textbf{4564} \\
              \midrule
              \parbox[t]{2mm}{\multirow{4}{*}{\rotatebox[origin=c]{90}{Synthetic}}} & Perfect.~rel.~& \textbf{166} & 92 & 40 & 150 & \multirow{4}{*}{0.26}\\
              & High.~rel.~& 227 & \textbf{188} & 197 & 305 \\
              & Related & 170 & 301 & \textbf{695} & 1871 \\
              & Irrelevant~& 25 & 66 & 168 & \textbf{3415} \\
            \bottomrule
        \end{tabular}
    \end{adjustbox}
\end{table}

Recent work \cite{thomas2023large} showed that it is possible to get very high-quality relevance judgements using LLMs. Thus, the question we investigate further is can we generate a fully synthetic test collection by not only generating synthetic queries but also by generating synthetic relevance judgements? To answer this question, we used the GPT-4 language model as accessed through Open AI's API in order to automatically label the documents (that were originally annotated using human annotators) for the synthetic queries to generate synthetic relevance judgements. For this purpose, we used the prompt template introduced in the recent study \cite{thomas2023large} that shows the possibility of getting high-quality relevance judgements using LLMs. In our experiments the temperature was set at zero, so the model would select the single most likely output and other parameters of the model were top $p$ = 1, frequency penalty 0.5, and presence penalty 0.

\begin{figure*}
    \centering
    \subfloat[Real vs.~Synthetic queries\label{fig:queries_realXgenerated}]
    {
        {
            \includegraphics[scale=0.35]{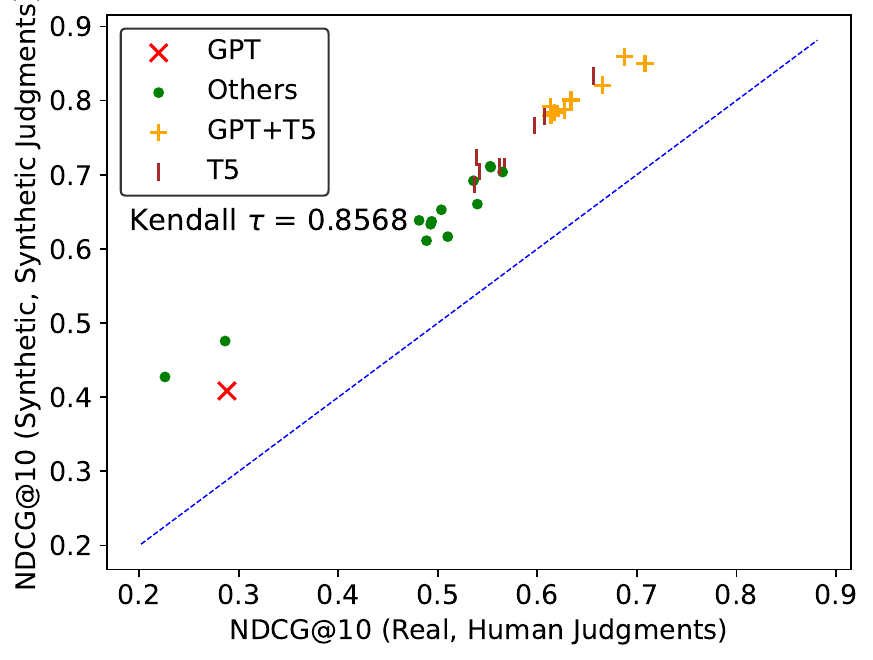}
        }
    }%
    % \quad
    \subfloat[Real vs.~GPT-4 queries\label{fig:queries_realXgpt4}]
    {{\includegraphics[scale=0.35]{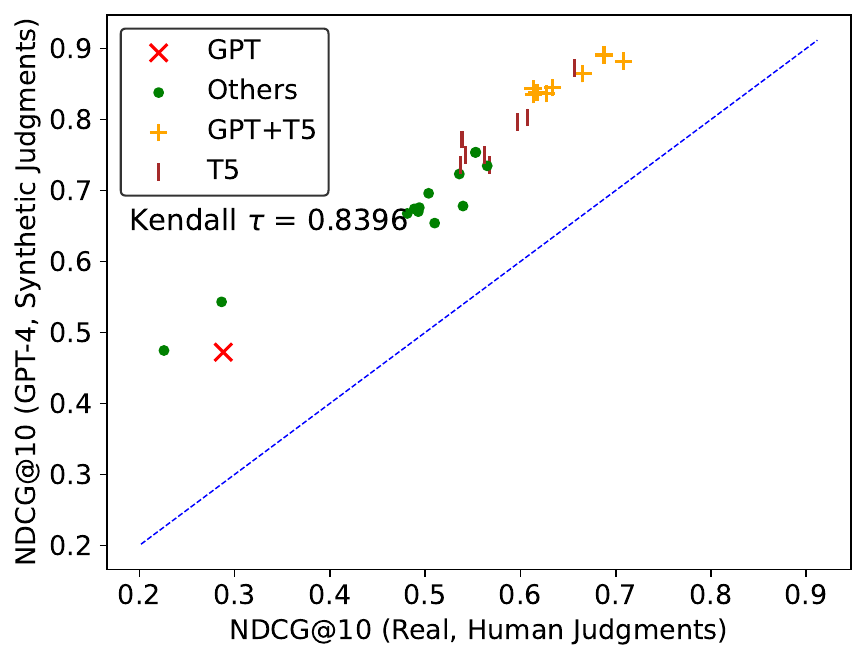} }}%
    % \quad
    \subfloat[Real vs.~T5 queries\label{fig:queries_realXt5}]
    {{\includegraphics[scale=0.35]{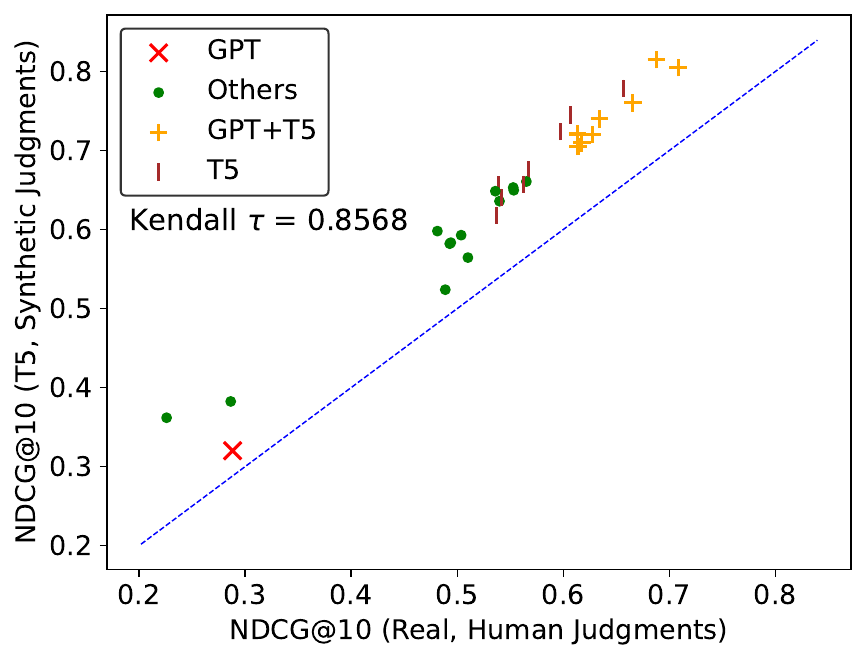} }}%
    \caption{Scatter plots of the effectiveness of TREC DL 2023 runs based on synthetic vs. real test collections to analyse the bias towards systems using the same language model as the one used in synthetic test collection construction.}%
    \label{fig:synthetic_query_nist}%
\end{figure*}

Table \ref{tbl:judgment_agreement_real_queries} shows how the judgments generated using GPT-4 compare with manual judgments. The table reports the ``agreement'' on the full 4-point relevance scale on real and synthetic queries, respectively. For both real and synthetic queries, we observe a fair level of agreement between synthetic judgements generated using GPT-4 and manual judgments: The Cohen's $\kappa$ on real queries is $0.24$ and on synthetic queries is $0.26$. In earlier studies, Faggioli et al.~\cite{faggioli2023perspectives} report a Cohen's $\kappa$ of $0.26$ between TREC DL 2021 assessors and GPT-3.5 with two types of human judgments, and Thomas et al.~\cite{thomas2023large} show that Cohen's $\kappa$ varies from $0.20$ to $0.64$ on two-level scale of TREC Robust data based on different versions of prompts. 

The generated judgments on synthetic queries show that GPT-4 consistently underestimates `Perfectly relevant' and `Highly relevant' labels. On real queries, GPT-4 labels on `Perfectly relevant' are more likely to be labelled as the same but on synthetic queries, according to GPT-4, a `Perfectly relevant' document is more likely to be rated as a `Highly relevant' document. For documents judged as `Perfectly relevant' by human assessors, GPT-4 generates the same judgment in 28\% of the cases. The results indicate that human assessors may use more detailed information to distinguish between `Perfectly relevant' from `Highly relevant' documents that are not fully captured by an LLM. We notice a similar pattern when we compare the `Highly relevant' and `Related' documents on both real and synthetic queries. 

One of the primary uses of test collections is for system evaluation. Even though the synthetic judgments generated could be slightly different than the actual judgments, they may still evaluate systems in a similar way. Figure \ref{fig:realXsynthetic_humanXllm} shows how the ordering of system on real queries compare with system ordering on our fully synthetic test collection (synthetic queries + synthetic judgments). It can be seen that evaluation on the fully synthetic test collection results in similar results to human queries with human judgments in terms of system ordering, with a Kendall's $\tau$ value of $0.8568$. For comparison, Faggioli et al.~\cite{faggioli2023perspectives} report Kendall's $\tau = 0.86$ for NDCG@10 on a similar experiment where only judgments were synthetically generated using GPT-3.5. Interestingly, compared with the system rankings using the human judgments on synthetic queries (see Figure \ref{fig:realXsynthetic_humanXhuman}), using the synthetic judgments generated with LLMs on synthetic queries result in higher correlations and Kendall's $\tau$ values as shown in Figure \ref{fig:realXsynthetic_humanXllm}.
\section{Analysis of Bias in System Evaluation}
\label{sec:bias-analysis}
One potential issue with using synthetic queries and judgements in test collection construction is the possible bias these collections may exhibit towards systems that are based on a similar approach (similar language model) to the one that was used in the synthetic test collection construction process (e.g., synthetic test collections constructed using T5 might favour systems that are based on T5). 

In order to analyse the possible bias, we categorised the runs submitted to TREC DL 2023 based on the approach they use\footnote{To this end, we carefully analysed the metadata file of submissions. GPT models are GPT-4 or GPT-3.5 and T5 models include MonoT5, FlanT5, and RankT5.} (i.e., language models used in their ranking or retrieval pipeline), resulting in four different system categories: systems based on GPT (\textcolor{red}{$\times$}), T5 (\textcolor{brown}{$|$}), GPT + T5 (\textcolor{yellow}{+}) (i.e., a combination of GPT and T5), and others ({\color{others}$\medbullet$}) (i.e., traditional methods such as BM25, or any model that does not use either GPT or T5). Figure~\ref{fig:queries_realXgenerated} shows that the synthetic test collection we have constructed that contains synthetic queries generated by LLMs (T5 and GPT-4) exhibits little to no bias towards LLM-based systems.

To further analyse possible bias that might arise from a system using a similar language model as the one used in test collection construction, Figure~\ref{fig:queries_realXgpt4} shows how system performance computed on synthetic test collections generated using queries generated by GPT-4 compare with system performance on real test collections. It can be seen that synthetic test collections based on GPT-4 do not systematically overestimate the performance of systems based on GPT. Similarly, as can be seen in Figure \ref{fig:queries_realXt5}, synthetic test collections with queries generated using T5 exhibit almost no bias towards systems based on T5. Similar results were obtained on other evaluation metrics such as average precision and NDCG@100, results for which are omitted due to space limitations. However, all three plots in the figure show that for all system types, systems consistently achieve higher performance on synthetic test collections when compared to real queries, suggesting that synthetic test collections tend to be easier than real queries and they tend to overestimate system performance across all system types. 

Although previous studies \cite{liu2023gpteval} on LLM evaluation discussed the potential bias towards LLM-generated text when we use LLMs for evaluation, in our experiments we did not observe a very clear evidence of systematic bias, where runs using GPT-4 were favored when evaluated using synthetic GPT-4 queries, or where runs using T5 were favored when evaluated on synthetic T5 queries. While our results look encouraging, we would like to emphasize that our results are based on one test collection we have constructed and further experiments are needed to analyse potential biases that might arise from using a fully synthetic test collection.
\section{Conclusion and Future Work}
\label{sec:conclusion}
In this paper, we explored the construction of fully synthetic test collections using LLMs. Overall, our analysis suggests that by using fully synthetic test collections consisting of synthetically generated queries and judgments, it is possible to obtain evaluation results that are similar to evaluation results obtained using the traditional test collection approach. Future studies can examine more advanced prompting methods and different LLMs to compare with the test collection we have created. 

Exploring the potential biases that may arise from generating a fully synthetic test collection is crucial for ensuring the quality, fairness, and reliability of the test collections. While our preliminary analysis showed that the synthetic test collections we have generated using a particular LLM do exhibit little to no bias towards systems based on the same LLM, further research is needed to provide a deeper understanding of the potential biases and to develop appropriate mitigation strategies.

\begin{acks}
This research is supported by the Engineering and Physical Sciences Research Council [EP/S021566/1] and the EPSRC Fellowship titled ``Task Based Information Retrieval'' [EP/P024289/1].
\end{acks}

\bibliographystyle{ACM-Reference-Format}
\bibliography{references}

% \appendix

\end{document}